\documentclass[11pt a4paper]{article}
\usepackage{jheppub}
\usepackage{bm}

\newcommand{\be}{\begin{equation}}
\newcommand{\dd}{\displaystyle}
\newcommand{\ee}{\end{equation}}
\newcommand{\bea}{\begin{eqnarray}}
\newcommand{\eea}{\end{eqnarray}}

\newcommand{\nn}{\nonumber}
\newcommand{\de}{\partial}
\def\nn{\nonumber}
\def\de{\partial}

 \def\slash#1{\setbox0=\hbox{$#1$}#1\hskip-\wd0\dimen0=5pt\advance
       \dimen0 by-\ht0\advance\dimen0 by\dp0\lower0.5\dimen0\hbox
         to\wd0{\hss\sl/\/\hss}}
\def\be{\begin{equation}}
\def\dd{\displaystyle}
\def\ee{\end{equation}}

\def\bea{\begin{eqnarray}}
\def\eea{\end{eqnarray}}

\def\7{\tilde}
\def\8{\hat}

 \def\slash#1{\setbox0=\hbox{$#1$}#1\hskip-\wd0\dimen0=5pt\advance
       \dimen0 by-\ht0\advance\dimen0 by\dp0\lower0.5\dimen0\hbox
         to\wd0{\hss\sl/\/\hss}}

\usepackage{color}
\usepackage{color}

\newcommand{\eq}[1]{(\ref{eq:#1})}
\newcommand{\ds}[1]{\frac {d#1}{ds}}

\begin{document}

\subheader{\hfill{\rm ICCUB-20-014}}

\title{A particle model with extra dimensions 
from Coadjoint Poincar\'e Symmetry}

\author[a]{Andrea Barducci}
\affiliation[a]{Department of Physics and Astronomy, University of Florence and
INFN, Via G. Sansone 1, 50019 Sesto Fiorentino (FI), Italy}

\author[a]{Roberto Casalbuoni}
\author[b]{Joaquim Gomis}
\affiliation[b]{Departament de F\'isica Qu\`antica i Astrof\'isica\\ and
Institut de Ci\`encies del Cosmos (ICCUB), Universitat de Barcelona\\ Mart\'i i Franqu\`es , ES-08028 Barcelona, Spain} 
\emailAdd{barducci@fi.infn.it}\emailAdd{casalbuoni@fi.infn.it}\emailAdd{gomis@ecm.ub.es} 
\keywords{Extra dimensions, point particle,  coadjoint Poincar\'e algebra}

\begin{abstract}
{Starting from the coadjoint Poincar\'e algebra we construct a point particle relativistic model with an interpretation in terms of extra-dimensional variables. The starting coadjoint Poincar\'e algebra is able to induce a mechanism of dimensional reduction between the usual coordinates of the Minkowski space and the extra-dimensional variables which turn out to form an antisymmetric tensor under the Lorentz group. 
 Analysing the dynamics of this model, we find that, in a particular limit, it is possible to integrate out the extra variables and determine their effect on the dynamics of the material point in the usual space time. 
 The model describes a particle
 in $D$ dimensions 
subject to a harmonic motion when one of the parameters
of the model is negative.
 The result can be interpreted as a modification to the flat Minkowski metric with  non trivial Riemann,  Ricci tensors  and scalar curvature  
 }

\end{abstract}
\maketitle
 \section{Introduction}

The space-time symmetry algebra of non-relativistic 
string theory \cite{Gomis:2000bd,Danielsson:2000gi} is the stringy Galilei algebra 
\cite{Brugues:2004an,Gomis:2005pg,Brugues:2006yd,Barducci:2018wuj}. This algebra contains  a vector  and  an antisymmetric two tensor
  "central" generators, and it  can be obtained
as a contraction of the coadjoint Poincar\'e algebra in $D$ dimensions 
\cite{Barducci:2019jhj} with generators
$P_\mu$ and $M_{\mu\nu}$ and  two abelian charges $Z_\mu$ and $Z_{\mu\nu}$. The numbers of generators of the algebra 
is $D(D+1)$.
 
 A dynamical  realization of the coadjoint Poincar\'e algebra in three dimensions has appeared  in \cite{Bergshoeff:2020fiz}
 where non-relativistic gravities
 in three dimensions
 \cite{Papageorgiou:2009zc,Bergshoeff-Rosseel,Obers-Hartong,
 Ozdemir:2019orp}
  have been obtained as a non-relativistic
 limit from the coadjoint  Poincar\'e gravity.

Here we will consider another realization of the
coadjoint Poincar\'e algebra in any dimension constructing
a particle action invariant under this
symmetry.  We will obtain the action using the non-linear
realization approach to space-time symmetries.


We will consider the Maurer-Cartan (MC) one-forms, constructed by the quotient of the coadjoint Poincar\'e algebra with  respect to the Lorentz group generators. The coset space is described by three types of coordinates, the Minkowski usual space-time coordinates, $x^\mu$ and the coordinates $\xi^{\mu\nu}$ and $\eta^\mu$ associated to the generators $Z_{\mu\nu}$ and $Z_\mu$ respectively. 
There are three MC forms associated to the generators in the coset and, in principle, this allows various kind of dynamical models,  describing particles and, more generally,  $p$-branes.

In this paper we will first concentrate on a particular particle model, using first the MC forms associated to the generators $P_\mu$ and $Z_{\mu\nu}$. We find that a combination of these forms describes a  free particle model in a space-time of $D+D(D-1)/2$ dimensions, parametrised by the coordinates $x^\mu$ and $\xi^{\mu\nu}$.
It is interesting to notice that, in this case, the extra-dimensions are not parametrised by Lorentz scalars but by an antisymmetric tensor. 

The signature of the Minkowski space time metric is mostly plus,
however the  total space time is a space with more that one time.
In $D=4$, the  signature is $(1,3)$. Instead,  the extended   space time has dimensions 10 and the signature $(4,6)$, since the
"electric" coordinates $\xi^{0i}$ have negative signature. 
 Of course, the model obtained in this way has a symmetry much larger that the original coadjoint Poincar\'e algebra. In fact, the symmetry group is the Poincar\'e group in $D(D+1)/2$ dimensions
whose generators  are $D(D+1)(D(D+1) -2)/ 8$, which is much larger than the generators of the coadjoint Poincar\'e algebra.
 
  The coadjoint Poincar\'e symmetry offers a simple way to do a dimensional reduction, through the MC one-form associated to the generator $Z_\mu$.
This form contains the coordinates $\eta^\mu$ and a coupling between the $x^\mu$'s and the $\xi^{\mu\nu}$'s. This coupling breaks the Poincar\'e symmetry in $D(D+1)/2$ space-time dimensions to the 
 coadjoint Poincar\'e symmetry in $D$ dimensions that contains
 the  Poincar\'e symmetry in the same number of space time dimensions.

The model depends crucially on two parameters, one, $z$ dimensionless, weighting the relevance of the extra dimensional variables $\xi^{\mu\nu}$ relatively to our space-time variables $x^\mu$. The other parameter $R$ with dimension of a length measures the relevance of the third MC form with respect to the first two. 
In other words, it measures the importance of the coupling between  $x^\mu$ and  $\xi^{\mu\nu}$. As we will show, the 
dynamics of the
variables  $\eta^\mu$ defines  a constant time-like vector, $F^\mu$,
which is the canonical momenta associated to $\eta^\mu$
written in terms of lagrangian variables. The dynamics 
allows to integrate the variables $\xi^{\mu\nu}$ in terms of 
$x^\mu$  in a particular limit where we send $R\to\infty$ and $z\to 0$ as $1/R^3$. Therefore we can construct an
effective action for $x^\mu$. This effective action breaks 
the coadjoint Poincar\'e symmetry 
 of
$D$ space time dimensions.
The effective action describes a particle in $D$ dimensions 
subject to a harmonic motion, for $z<0$,  on a $(D-1)$
subspace orthogonal to the constant time-like vector $F^\mu$. 
This gives rise to a modification of the mass of the particle which, in the first quantised version produces a tower of particle with mass spaced by the frequency of the oscillator, $\omega\sim 1/\sqrt{R}$.
 We show that this situation can be also described as a modification to the flat Minkowski metric 
 in D dimensions
 by a quadratic term in the coordinates transverse to the vector $F^\mu$. The modification of the flat metric is non trivial, giving rise to non zero Riemann,  Ricci tensor  and scalar curvature.  
There are other two limits, one sending   $z\to 0$ with $R$ fixed. In  this case we obtain   a free particle in $D$ dimensions (see later). 
The other is  $R\to\infty$ with fixed $z$  which  describes a free particle in $D(D+1)/2$ dimensions.


The paper is organised in the following way: in Section 2 we introduce the coadjoint Poincar\'e algebra and we evaluate the three MC forms.
 In Section 3 we discuss a model  which can be interpreted as describing a free particle in $D(D+1)/2$ dimensions and  we analyse  its symmetries. In Section 4 we consider  the dimensional reduction of the previous model through the addition of a second term to the action obtained from the third MC form. The resulting model  is invariant under two distinct diffeomorphisms (diff).
In Section 4.1 we discuss the constraints induced by the the diff-invariance. In Section 4.2 we consider the limit where $z\to 0$ and $R\to\infty$ in a correlated way and proceed to integrate out the variables $\xi^{\mu\nu}$, obtaining an effective action. 
In Section 5 we summarise the main results obtained in this paper and we discuss also some perspective as, for instance, the extension to strings and $p$-branes.

\section{Maurer-Cartan one-forms}\label{sec:I}

The  coadjoint Poincar\'e algebra 
 is an extension of the Poincar\'e algebra with a  vector and an antisymmetric rank-two tensor, $Z_\mu$ and $Z_{\mu\nu}$ respectively, satisfying the commutation relations 
 \cite{Barducci:2019jhj}
 \bea
  \left[M_{\mu\nu},Z_{\rho}\right]&=&i(\eta_{\mu\rho}Z_{\nu}-\eta_{\nu\rho}Z_{\mu}),
\nn\\
 \left[M_{\mu\nu},Z_{\rho\sigma}\right]&=&i(\eta_{\mu\rho}Z_{\nu\sigma}+\eta_{\nu\sigma}Z_{\mu\rho}
 -\eta_{\mu\sigma}Z_{\nu\rho}
  -\eta_{\nu\rho}Z_{\mu\sigma})\nn\\
    \left[Z_{\mu\nu},P_{\rho}\right]&=&i(\eta_{\mu\rho}Z_{\nu}-\eta_{\nu\rho}Z_{\mu}),
\nn\\
  \left[Z_{\mu},P_{\nu}\right]&=&0 ,~~~  \left[Z_{\mu},Z_{\nu}\right]=0,~~~
,~~~ \left[Z_{\mu\nu},Z_{\rho}\right]=0, ~~~
\mu, \nu =0,1\cdots D-1
  \label{eq:1.1}
\eea
Let us consider the quotient space of the group generated by this algebra with respect to the Lorentz group.  A local parameterisation of the
coset is given by 
\be
g=e^{ix^\mu P_\mu} e^{\frac i2\xi^{\mu\nu} Z_{\mu\nu}}e^{i\eta^\mu Z_\mu}
\label{eq:1.2}\ee
The Maurer Cartan (MC) one-form is given by
\be
\Omega=-ig^{-1}dg= dx^\mu P_\mu+(d\eta^\mu-\xi^{\mu\nu} dx_\nu)Z_\mu+\frac 12 d\xi^{\mu\nu} Z_{\mu\nu}\label{eq:1.3}
\ee
or:
\be
\Omega= \Omega_1^\mu P_\mu+\frac 12\Omega_2^{\mu\nu} Z_{\mu\nu}+\Omega_3^\mu Z_\mu
\ee
where
\be
\Omega_1^\mu = dx^\mu,~~~\Omega_2^{\mu\nu} =d\xi^{\mu\nu},~~~\Omega_3^\mu=d\eta^\mu-\xi^{\mu\nu} dx_\nu.
\label{eq:1.4}
\ee

It is easy to verify that the MC forms are invariant under the following transformations:
\bea\label{coadjoint_tras}
P_\mu:~~~\delta x^\mu&& =a^\mu\nn\\
Z_{\mu\nu}: \delta \xi^{\mu\nu} &&=\epsilon^{\mu\nu}, ~~~\delta\eta^\mu =\epsilon^{\mu\nu} x_\nu\nn\\
Z_\mu:~~~\delta\eta^\mu&& =\epsilon^\mu\label{eq:1.13}
\eea
These transformations are generated by  the right invariant vector fields
\be
P_\mu =-i\frac\de{\de x^\mu},~~~Z_\mu =-i\frac\de{\de\eta_\mu},~~~
Z_{\mu\nu}= -i\frac\de{\de\xi^{\mu\nu}}+i \left(x_\mu \frac\de{\de\eta^\nu}-x_\nu \frac\de{\de\eta^\mu}\right))\label{eq:17}
\ee
 Furthermore the Lorentz group generators are
\be
M_{\mu\nu}= -i\left(x_\mu \frac{\de}{\de x^\nu}-x_\nu \frac{\de}{\de x^\mu}\right)
-i\left(
\eta_\mu \frac{\de}{\de \eta^\nu}-\eta_\nu \frac{\de}{\de \eta^\mu}\right)-\ i 
\left(\xi_{\mu\rho}\frac{\de}{\de\xi_\rho^{\,.\nu} }- \xi_{\nu\rho}\frac{\de}{\de\xi_\rho^{\,.\mu}}
 \right)\label{eq:16}\ee

\section{A model with extra-dimensions}
\def\ein{\dd{\sqrt{-{ \dot x_\mu^2-\frac{z}{2}\dot\xi_{\mu\nu}^2}}}}
\def\einc{\sqrt{-(\dot \eta^\mu-\xi^{\mu\nu} \dot x_\nu)^2}}
Starting from the MC forms given in \eq{1.4} 
and using the non-linear
realization approach to space-time symmetries,
one can construct many actions invariant under the coadjoint Poincar\'e algebra. 
Here we will consider an
action involving only the 
coordinates $x^\mu,\, \xi^{\mu\nu}$ through the pullback of 
two MC forms $\Omega^\nu_1$ and $\Omega^{\mu\nu}_2$:
\be
S_1=-M\int d\tau\,\ein\,,\label{eq:556}
\ee
 The sign of the parameter $z$  is simply related to the sign with which the "electric" , $\xi^{0i}$, and  the "magnetic components, $\xi^{ij}$, contribute to the line element in \eq{556}.

We assume also the following dimensions in mass:
\be
[x^\mu\ ]= [\xi^{\mu\nu}]= -1,
\ee

Let us nos consider the global symmetries of this lagrangian. Since the $\eta_\mu$ variables do not appear, we can eliminate all this dependence on the generators of eqs. \eq{17} and \eq{16}. It follows that, for the moment being,  the symmetries generated by $Z_\mu$ can be simply ignored.
Also, the variables $x_\mu$ and $\xi_{\mu\nu}$ 
are decoupled. Therefore, the model is invariant under  two independent Lorentz groups, one acting on the position variables and the other on $\xi_{\mu\nu}$. It is convenient to define the following quantities:
\be
M^1_{\mu\nu}= -i\left(x_\mu \frac{\de}{\de x^\nu}-x_\nu \frac{\de}{\de x^\mu}\right)
\ee
and
\be
M^2_{\mu\rho,\sigma\nu}=-\frac i 2
\left(\xi_{\mu\rho}\frac{\de}{\de\xi^{\sigma\nu} }- \xi_{\sigma\nu}\frac{\de}{\de\xi^{\mu\rho}}
 \right)
\ee
Notice that the part of the original Lorentz group generators, excluding the $\eta_\mu$ part, is given by
\be
M^1_{\mu\nu} +2\eta^{\rho\sigma}M^2_{\mu\rho,\sigma\nu}
\ee
We see that $L_1$ is also invariant under the translations in $x_\mu$ and $\xi_{\mu\nu}$ generated by $P_\mu$ and $Z_{\mu\nu}$ respectively. That is to say, our lagrangian is invariant under the two Poincar\'e groups acting on $x^\mu$ and $\xi^{\mu\nu}$. 
In 4 dimensions we have  $ 10+21=31$ symmetries.

.However the full symmetry group of $L_1$ is much larger.   In fact 
introducing the following  variables 
\be
y^A=( x_\mu, \sqrt{\frac{|z|}2}\xi^{\mu\nu})
\ee
we can write $L_1$ as a lagrangian for a free particle in $D(D+1)/2$ dimensions:
\be
L_1=-M\sqrt{- \eta_{\mu\nu} \dot x^\mu\dot x^\nu-\frac{z}2\eta_{\mu\rho}\eta_{\nu\sigma}\dot\xi^{\mu\nu}\dot\xi^{\rho\sigma}}=-M\sqrt{-\eta_{AB}\dot y^A\dot y^B}
\ee
with
\be
\eta_{AB} =\left(\eta_{\mu\nu}, \epsilon (z)\eta_{\mu\rho}\eta_{\nu\sigma}\right)
\ee
 In the particular case of $D=4$ this is the lagrangian of a free particle moving in 10 dimensions.

 The symmetry  algebra of $L_1$ is the Poincar\'e algebra in $D+D(D-1)/2$ dimensions.  For example, for $D=4$, the Poincar\'e group in 10 dimensions has $55$ generators. In general, the generators of this enlarged Poincar\'e group are given by
\be
P_A=-i\frac{\de}{\de y^A}, ~~~S_{AB}=-i\left(\eta_{AC}\, y^C\frac{\de}{\de y^B}- \eta_{BC}\,y^C\frac{\de}{\de y^A}\right)
\ee
Therefore, besides the two Poincar\'e symmetries there are other symmetries intertwining the space of the $x^\mu$ with the one of the one spanned by $\xi^{\mu\nu}$:
\be
S_{\mu\nu,\rho}=- i\left(\sqrt{\frac{|z|}2}(\xi_{\mu\nu}\frac{\de}{\de x^\rho}-\sqrt{\frac 2{|z|}}x_\rho\frac{\de}{\de\xi_{\mu\nu}}\right)
\ee
which together with the Poincar\'e generators  in $D$ and $D(D-1)/2$ dimensions span the full algebra of the Poincar\'e group in $D(D+1)/2$ dimensions. This analysis corresponds to the following decomposition  of the Poincar\'e algebra in $D(D+1)/2$ dimensions:
\vskip0.5cm
\begin{center}
 Poincar\'e in  $D$  dim $\oplus$   Poincar\'e in $ \frac{D (D-1)}2$  dim $\oplus  S_{\mu\nu,\rho}$
 \end{center}
 \vskip0.5cm
As said before, the generators $S_{\mu\nu,\rho}$ intertwine the $x^\mu$ with the $\xi^{\mu\nu}$-space. Or, said in other way,  they intertwine  two different representations of the original Lorentz group in $D$ dimensions, namely: $(1/2,1/2)$ and $(1,0)\oplus(0,1)$.

The analysis of the generators of the large Poincar\'e algebra in terms of the original Lorentz algebra in  \eq{16} is made explicit in the tables \ref{Tab1} and \ref{Tab2}. Besides giving the details for $D=4$ we do the same for $D=3$. This case is of particular interest because all the dynamical variables are three-vectors: $x_\mu, \xi_\mu=\frac 12 \epsilon_{\mu\nu\rho}\xi^{\mu\nu}, \eta_\mu$.

\begin{table}[h]
\caption{Translation generators}
\begin{center}
\begin{tabular}{||c|c|c|c|||}
\hline\hline
generators&dimensions &~ D=3 ~~&~ D=4~~\\
\hline\hline
$P_A$ & $D(D+1)/2$& 6& 10\\
\hline\hline
$P_\mu$ & $D$ & 3& 4\\
\hline
$Z_{\mu\nu}$ & $D(D-1)/2$& 3& 6\\
\hline\hline
\end{tabular}
\end{center}
\label{Tab1}
\end{table}

\begin{table}[h]
\caption{Lorentz generators}
\begin{center}
\begin{tabular}{||c|c|c|c||}
\hline\hline
generators&dimensions &~ $D=3$~~&~ $D=4$~~\\
\hline\hline
$S_{AB}$ & $D(D+1)(D(D+1) -2)/ 8$ &15 &45 \\
\hline\hline
$S_{\mu\nu}$ & $D(D-1)/2$ & 3 &6\\
\hline
$S_{\mu\nu;\rho\sigma}$ & $D(D-1)(D(D-1)-2)/8$ & 3 &15\\
\hline
$S_{\mu\nu;\rho}$ &$D( D(D-1)/2$& 9 &24\\
\hline\hline
\end{tabular}
\end{center}
\label{Tab2}
\end{table}


\section{Dimensional reduction}

We will now introduce a further term in the action, 
Using the  pull back of the Maurer-Cartan one-forms,
 of Section \ref{sec:I},
\be
S=\int d\tau (L_1+L_2)=S_1+S_2
\label{eq:27}\ee
where
\be
L_1=-M\sqrt{-{\dot x_\mu^2-\frac{z}{2}\dot\xi_{\mu\nu}^2}}\,~~~L_2=- \frac 1{R^2}\sqrt{-(\dot \eta^\mu-\xi^{\mu\nu} \dot x_\nu)^2}
\ee
The model contains an extra vector $ \eta_\mu$ 
associated to the generator $Z_\mu$.
Given the dimensions in mass of $x^\mu$ and $\xi^{\mu\nu}$, the dimension in mass of $\eta^\mu$ is given by:
\be
[\eta_\mu] = -2
\ee

An interesting feature appears due the introduction of the term depending on the MC form $\Omega_3^\mu$:
$\einc$. The Poincar\'e algebra in $D(D+1)/2$,
with many times, 
is explicitly broken to the Poincar\'e algebra in $D$ dimensions, but there is an emergent symmetry  generated by $Z_\mu$ which was acting trivially on the lagrangian \eq{556}, 
Therefore this term gives rise automatically to a dimensional reduction from $D(D+1)/2$  to $D$ dimensions.
It should be noticed that the extra-dimensional variables $\xi^{\mu\nu}$, at difference with the usual models with extra dimensions, are not Lorentz scalars, but rather they  belong to a non-trivial representation of the Lorentz group.

The action \eq{27}, besides the global invariance under the coadjoint Poincar\'e algebra,  it is invariant  under two diffeomorphisms, one associated to $S_1$ and the other to $S_2$. To study this point, let us consider 
the  Euler-Lagrange derivatives of the
total action 
\bea
\frac{\delta S}{\delta x^\mu}
&=&-\frac {d}{d\tau} \Big(M\frac  {\dot x_\mu}{\ein}
-F_\rho\xi^\rho_{\,.\mu}\Big) \label{eq:43}\\
\frac{\delta S}{\delta \xi^{\mu\nu}}
&=& -\frac d{d\tau}\Big(zM\frac{\dot \xi_{\mu\nu}}{\ein}\Big)-\Big(
F_\mu\dot x_\nu-F_\nu\dot x_\mu\Big)\label{eq:44}\\
\frac{\delta S}{\delta \eta^\mu}& =&-\frac d{d\tau}F_\mu\label{eq:45}\eea
where 
\be
F_\mu=\frac 1{R^2}\frac  {(\dot \eta_\mu-\xi_{\mu\rho} \dot x^\rho)}{\einc}\label{eq:29}
\ee

The Euler-Lagrange derivatives are invariant under 
coadjoint Poincar\'e transformations (\ref{coadjoint_tras})
 and  are not independent. In fact, we have  two Noether identities (see for example \cite{Batalin:1985qj, Henneaux:1992ig,Gomis:1994he}), the first 
 one associated to the diff-invariance of the total action, and the second one to the diff-invariance of $S_2$:
 \bea
&&\dot x^\mu\frac{\delta S}{\delta x^\mu}+\frac 12 \dot\xi^{\mu\nu} \frac{\delta S}
{\delta \xi^{\mu\nu}}\label{eq:46}
+\dot\eta^\mu\frac{\delta S}{\delta \eta^\mu}=0\\
&&(\dot \eta^\mu-\xi^{\mu\rho}{\dot x_{\rho}})\frac{\delta S}{\delta \eta^\mu}=0.\label{eq:47}
\eea
Therefore we have two gauge symmetries
\bea\label{gauge1}
&\delta_1 x_\mu= \rho_1\dot x_\mu&\nn\\
&\delta_1 \xi_{\mu\nu}=\rho_1\,\dot\xi_{\mu\nu}&\nn\\
&\delta_1\eta_\mu= \rho_1\dot\eta_\mu&\nn\\
&\delta_2 x_\mu=\delta_2 \xi_{\mu\nu}=0,~~~\delta_2\eta_\mu=\rho_2\,(\dot \eta_\mu-\xi_{\mu\rho} \dot x^\rho)&\label{eq:449}
\eea
where $\rho_1$ and $\rho_2$ are arbitrary functions of $\tau$.
The transformation with parameter $\rho_1$ is the ordinary world-line diffeomorphism. We can check explicitly that these transformations leave the lagrangian 
(\ref{eq:27}) invariant
 up to a total derivative:
\be
\delta_1 L=\frac d{d\tau}({\rho_1 L})
\ee
and
\be
\delta_2 L_1=0,~~~\delta_2 L_2=c\frac d{d\tau}({\rho_2 L_2})
\ee
It is convenient  define   the analogue of the proper time for this model as:
\be
ds = \ein d\tau\label{eq:552}
\ee
Notice, that this is nothing but the proper time for the particle in $D(D+1)/2$ dimensions. Of course, the choice of this parameter is equivalent to the  gauge choice: 
\be
\ein=1
\ee
Using the expressions \eq{43}-\eq{45} and \eq{46} inside \eq{47} we get the identity:
\be
\frac{d x^\mu}{ds} L^1_\mu +\frac 12 \frac{d\xi^{\mu\nu}}{ds} L^2_{\mu\nu}\equiv  0\label{eq:52}
\ee
where
\be
L^1_\mu=M \frac d{ds}  \ds{x_\mu}-F_\rho \ds{\xi^{\rho}_{\,.\mu}}
\ee
\be
L^2_{\mu\nu}=zM \frac d{ds}\ds{ \xi_{\mu\nu}}+
\left(F_\mu\ds{ x_\nu}-F_\nu\ds{ x_\mu}\right)\label{eq:39}
\ee

 The equations of motion are obtained from the vanishing of the expressions 
 \eq{43}, \eq{44} and \eq{45}. Using  $\dot F_\mu=0$ (from \eq{45}) in eq. \eq{43}, the other two equations of motion  can  be written
 as
 \be
 L^1_\mu=0,~~~L^2_{\mu\nu}=0
 \ee
 The identity \eq{52} shows that the two equations of motion are not independent one from the other.
 
 There are two interesting limiting situations that we could consider:\\
\noindent
1) - $R\to\infty$ with $z$  fixed and different from zero. In this case the action \eq{27} reduces to the first term, that is the action of a free particle in $D(D+1)/2$ dimensions.\\
\noindent
\be
S=-\int d\tau M\sqrt{-\dot x_\mu^2}
\ee
2) -  $z\to 0$ with fixed $R$. This would correspond to the choice
\be
S = \int\,d\tau\left(- M\sqrt{\dot x_\mu^2}- \frac 1{R^2}\sqrt{-(\dot \eta^\mu-\xi^{\mu\nu} \dot x_\nu)^2}\right)
\ee

 Then, the kinetic term for $\xi^{\mu\nu}$ vanishes and these variables become non-dynamical. From their variations we get

\be
\left(F_\mu\ds{ x_\nu}-F_\nu\ds{ x_\mu}\right)=0,
\ee
which has the solution
\be
F_\mu=f(\tau)\dot x_\mu
\ee
Then, from the condition that $F_\mu$ is a time-like vector,  $F^2=-1/R^4$, we get
\be
F_\mu=\frac 1{R^2} \frac {\dot x_\mu}{\sqrt{-\dot x^2}}
\ee
But  $F_\mu$ is constant, therefore
\be
\frac d{d\tau} \frac {\dot x_\mu}{\sqrt{-\dot x^2}}=0
\ee
which it is the equation of motion of a free particle.

 A third possibilities is to send $R\to \infty$ and $z\to 0$ in a correlated way, namely $z\approx 1/R^3$.
 This case will be discussed in detail in Section 4.2

\subsection{Hamiltonian analysis and canonical action}

As we have seen, the model described in this Section admits two gauge symmetries. Therefore we expect the presence of two 
 first class constraints
in the phase space. To this end, let us perform the hamiltonian analysis  of (\ref{eq:27}).

We start by computing the canonical momenta
\be
p_\mu= {M} \frac{\dot x_\mu}{\ein} -\frac 1 {R^2}\frac {(\dot \eta_\nu-\xi_{\nu\rho} \dot x^\rho)\xi^\nu_{.\,\mu}}{\einc}
\ee
\be
 \pi_\mu=\frac 1{R^2}\frac  {(\dot \eta_\mu-\xi_{\mu\rho} \dot x^\rho)}{\einc}\equiv F_\mu
\ee
\be
 \pi_{\mu\nu}=zM\frac{\dot\xi_{\mu\nu}}{\ein}
\ee
where we have introduced the quantity $F_\mu$, as  the momentum associated to $\eta^\mu$ in 
terms of the lagrangian variables.

From the previous expressions we see that there are two
 primary constraints
\be
\phi_1=\frac 12\left( (p_\mu+\pi_\rho\xi^\rho_{\,.\mu})^2+ \frac 1 {2z} \pi_{\mu\nu}^2  +M^2\right) =0,~~~      \phi_2=\frac 12(\pi_\mu^2+\frac 1{R^4})=0
\ee

These two constraints are first class:
\be
\{\phi_1,\phi_2\}=0.
\ee
 Therefore there are no secondary constraints. Notice that in the two limits that we have previously considered, only the constraint $\phi_1$ survives. In the first case we get the mass-shell constraint for a particle in $D(D+1)/2$ dimensions, whereas, in the second case we get the mass-shell condition for a particle in $D$ dimensions.

 The presence of two first class constraints implies the existence of two gauge transformations given by
\be
\bar\delta_iA=\{A, \epsilon_i\phi_i\}, 
\ee
where $\epsilon_i(\tau) $ are the gauge parameters. These transformations are the same as the ones  given in \eq{449} after
the following identification of the parameters $\rho_1,\rho_2$ 
\be
\epsilon_1=\frac 1 M\rho_1 {\ein},\quad
\epsilon_2=R^2 \rho_2{\einc} 
\ee
with the exclusion of  $\bar\delta_1\eta_\mu$ for which the  variation is a combination of two of the gauge transformations of \eq{449}:
\be
\bar\delta_1\eta_\mu= \rho_1\xi_{\mu\rho}\dot x^\rho= \delta_1\eta_\mu-\frac{\rho_1}{\rho_2}\delta_2\eta_\mu
\ee
Under these transformations we have
\be
\bar\delta_1 L_1=M\frac d{d\tau}({\epsilon_1 L_1}),~~~\bar\delta_1 L_2=0
\ee
and
\be
\bar\delta_2 L_1=0,~~~\bar\delta_2 L_2=\frac 1{R^2}\frac d{d\tau}({\epsilon_2 L_2})
\ee

The canonical action is given by 
\be
S_c=\int d\tau\big(p_\mu\dot x^\mu+\frac 1 {2} \pi_{\mu\nu}\dot\xi^{\mu\nu}+\pi_\mu\dot\eta^\mu
-e_1\phi_1-e_2\phi_2\big)
\ee
 The degrees of freedom in phase space are
$2D$ for 
$x^\mu , p_\mu $, $D(D-1)$ for $\xi^{\mu\nu},
\pi_{\mu\nu}$ and $2D$ for $\eta^\mu,\pi_\mu$. The physical
degrees of freedom will be 
$2D+D(D-1)+2D-2\times 2=D^2+
3D-4$, where we used the fact that there are two first  class
constraints.

\subsection{The effective action}

In this Section we would like to consider the limit $R\to\infty$ and $z\to 0 $  as $z\to 1/ R^\gamma, \gamma>0$.
 We will integrate out the variables $\xi^{\mu\nu}$ at the first  non vanishing order in  $1/R$.
 obtaining an effective action which, at the lowest order  describes a free particle in $D$ dimensions, whereas at the next order the equations of motion describe a particle in a quadratic potential, that could be also interpreted as a correction to the Minkowski flat metric. The quadratic potential involves only the coordinates orthogonal to the time-like constant vector $F^\mu$ defined in \eq{29}.
 This effective action breaks 
the coadjoint Poincar\'e symmetry in $D$ space time dimensions.
The effective action describes a particle in $D$ dimensions 
subject to a harmonic motion, for $z<0$.

In order to obtain an effective action from the decoupling of the extra dimensions, we will consider the equations of motion for $x^\mu$ and $\xi^{\mu\nu}$ in an arbitrary gauge 
and using $\dot F^\mu =0$
\be
\frac {d}{d\tau} \Big(M\frac  {\dot x_\mu}{\ein}\Big)=
F_\rho\dot\xi^\rho{}_{\mu} \label{eq:66}
\ee
\be
 \frac d{d\tau}\Big(zM\frac{\dot \xi_{\mu\nu}}{\ein}\Big)=-\Big(
F_\mu\dot x_\nu-F_\nu\dot x_\mu\Big)\label{eq:67}
\ee
Let us introduce the coordinates longitudinal and transverse
to $F^\mu$,
\be
x^L=\frac{F\cdot x}{\sqrt{-F^2}}, ~~~x^T_\mu= x_\mu-F_\mu\frac{F\cdot x}{F^2},
\ee
implying $x^2=-(x^{L})^2+(x^T)^2$.
Making use of $ \dot F^\mu =0$,we integrate  eq. \eq{67} obtaining
\be
zM\frac{\dot \xi_{\mu\nu}}{\ein}=-\Big(
F_\mu x_\nu^T-F_\nu x_\mu^T+\alpha_{\mu\nu}\Big)\label{eq:59}
\ee
where
\be
\alpha_{\mu\nu}= -zM\frac{\dot \xi_{\mu\nu}}{\ein}\Big|_{\tau=0}-\Big(
F_\mu x_\nu^T(0)-F_\nu x_\mu^T(0)\Big)
\ee
In order to integrate out completely the dependence on the $\xi$ variables, we need to eliminate in $\alpha_{\mu\nu}$ the dependence on the initial condition on $\dot\xi_{\mu\nu}$.This can only be done assuming that  the first term in the expression of $\alpha_{\mu\nu}$ vanishes in the limit we are considering. Therefore, we will require that for $R\to\infty$, the parameter $z$ goes to zero faster than the second term, that is: $z\to 1/R^3$. 
Then the eq. \eq{59} can be written as:
\be
zM\frac{\dot \xi_{\mu\nu}}{\sqrt{-\dot y^2-\frac z2\dot\xi^2}}=-\Big(
F_\mu y_\nu^T-F_\nu y_\mu^T\Big)\label{eq:70}
\ee
with
\be 
y_\mu=x_\mu-x_\mu(0)
\ee
Squaring the eq. \eq{70} we get
\be
\dot\xi^2= -\frac{T^2}{z^2 M^2}\,\frac{\dot y^2}{1+T^2/(2zM^2)}
\ee
where 
\be
T_{\mu\nu}=-\Big(
F_\mu y_\nu^T-F_\nu y_\mu^T\Big)
\ee
from which
\be
\dot x^2+\frac z2\dot \xi^2=\dot y^2+\frac z2\dot \xi^2=\frac1{1+T^2/(2zM^2)}\dot y^2
\ee

It is convenient to define
\be
W(y^T) =\frac{1}{1+T^2/(2zM^2)}=\frac{1}{1-(y^T)^2/(zM^2R^4)}\label{eq:63}
\ee

Substituting into  the equation of motion for $x_\mu$, \eq{66},  we get
\be
\frac 1{\sqrt{-W\dot y^2}}\frac d{d\tau} \frac {\dot y_\mu}{\sqrt{-W\dot y^2}}- \frac 1{zM^2 R^4}y_\mu^T=0\label{eq:75}
\ee
 
 If we introduce  the gauge fixing $\sqrt{-W\dot y^2}=1$,
 the equations of motion on this gauge become
 \be
 \ddot y^\mu-\frac 1{zM^2 R^4}y_\mu^T=0
 \ee
 which for $z<0$ represents a harmonic motion in the transverse coordinates.

 The diff invariant equation of motion (\ref{eq:75})
 can be obtained from the action
\be
S=-M\int d\tau\,\sqrt{-\left( 1-\frac{(y^T)^2}{zM^2R^4}\right)\dot y^2}\label{eq:65}
\ee

Notice that the action \eq{65} can be interpreted as the action of a particle in a background metric given by
\be
g_{\mu\nu}=\left(1-\frac{(y^T)^2}{zM^2R^4}\right)\eta_{\mu\nu}
\ee
The Christoffel symbols from this metric are given by
\be
\Gamma_{\mu\nu}^\sigma=-\frac 1{zM^2R^4}\left(y_\mu^T\delta_\nu^\sigma +y_\nu^T \delta_\mu^\sigma-(y^T)^\sigma\eta_{\mu\nu}  \right)
\ee
In order to evaluate the Riemann tensor we notice that the quadratic term in the Christoffel symbols is of higher order in $1/R$ and therefore  can be neglected:
\be
R^\sigma_{\rho\mu\nu}\approx\de_\nu\Gamma^\sigma_{\mu\rho}-\de_\mu\Gamma^\sigma_{\nu\rho}=
-\frac 1{zM^2R^4}\left(P_{\nu\rho}\delta^\sigma_\mu-P_{\mu\rho}\delta^\sigma_\nu +P^\sigma_\mu\eta_{\nu\rho} -                P^\sigma_\nu\eta_{\mu\rho}\right)
\ee

where 
\be
P_{\mu\nu}=\eta_{\mu\nu}-\frac{F_\mu F_\nu}{F^2}
\ee
Then, we have

\be
R_{\mu\nu}=R^\rho_{\mu\rho\nu}=-\frac 1{zM^2R^4}\left( (D-1)\eta_{\mu\nu}+ (D-2)P_{\mu\nu}          \right)
\ee
and, for the scalar curvature:
\be
R_\mu^\mu=-\frac 2{zM^2R^4}(D-1)^2
\ee
In the particular reference frame $F^\mu=(1/R^2,\vec 0\,)$, we get
\be
R_{00}=+\frac 1{zM^2R^4}(D-1),~~~R_{ij} =-\frac {(2D-3)}{zM^2R^4}\delta_{ij}
\ee

Notice also that  the square  mass term associated to the extra variables in the original formulation is $zM^2$. Therefore this gravitational field is entirely  dependent on  the extra dimensional space.

As we have shown at the lowest order in $1/R$ we get a particle moving in a quadratic potential in the transverse variables $y^T$. However, the action depends on a constant vector $F_\mu$ that ,at this level, should be thought as a given vector. Correspondingly the Lorentz invariance and  the  spatial translations are broken. On the other hand the action is invariant under a diffeomorphism, therefore we expect a first class constraint. In fact, from
\be
p_\mu=M\sqrt{1-\frac{(y^T)^2}{zM^2R^4}} \frac {\dot y_\mu}{\sqrt{-\dot y^2}}
\ee
we get
\be
p^2+M^2\left(1-\frac{(y^T)^2}{zM^2R^4} \right)
  =0\label{eq:76}
\ee
At the lowest order we can evaluate the shift in mass due to the perturbation originating from the extra coordinates; let us put
\be
p^L= M+\epsilon
\ee
Inserting this expression in the mass shell condition we get
\be
\epsilon = \frac 1{2M} (p^T)^2- \frac 1{2zM R^4}(y^T)^2
\ee
Therefore, if $z<0$, from the quantum point of view this means that there are no   negative energy states, and the shift in mass is given by
\be
\epsilon = \hbar\omega(n +1/2)\label{eq:79}
\ee
with
\be
\omega^2= \frac 1{|z| M^2R^4}
\ee

Notice that we could have started from the Klein-Gordon equation associated to the mass-shell constraint \eq{76} and deriving the energy eigenvalues. The result would have been $E_n^2=M^2+2M\hbar \omega(n+1/2)$. Expanding the positive energy solution at the order $1/R$, we find that the correction to $p^L$ is the one in \eq{79}.

 Summing up we have seen that starting from the original action
\eq{27},   invariant under coadjoint Poincar\'e transformations,  and integrating out  the variables
$\xi^{\mu\nu}$  (the variables associated to the extra 
dimensions), the associated effective action \eq{65}
describes for $z<0$ a  harmonic oscillator.

\section{Conclusions and outlook}

In this paper we have considered a dynamical model based on a non-linear representation of the coadjoint Poincar\'e group, which has a natural interpretation in terms of extra dimensions. At difference with the models of this sort, here the extra dimensions are described by an antisymmetric tensor with respect to the Lorentz group and contains many times.

Given our space-time in $D$ dimensions, the total space has $D(D+1)/2$ dimension, meaning 10 dimensions for $D=4$. The dimensional reduction from $D(D+1)/2$ dimensions to $D$ is performed through a term coupling together the coordinates $x^\mu$ and $\xi^{\mu\nu}$, This term introduces another vector $\eta^\mu$ giving rise to a constant time-like vector, $F^\mu$ (the momentum conjugated to $\eta^\mu$), but always maintaining the invariance under the coadjoint Poincar\' e symmetry.

 We have studied the model in a particular limit, where it is possible to integrate out the extra dimensional coordinates. The effective action obtained in this way describes a particle in $D$ dimension moving in a quadratic potential in the $D-1$ space orthogonal to the time-like vector $F^\mu$. In the case of z negative, in a first quantised version, this gives rise to a mass spectrum of a harmonic oscillator. The quadratic potential has an interesting interpretation in terms of a modification to the Minkowski flat metrics. The resulting metrics depends on the constant  time-like vector $F^\mu$  and has non-vanishing Riemann, Ricci tensor  and scalar curvature. 
 
 In practice we have shown that the existence of extra dimensions, with non trivial behaviour with respect to the Lorentz group, is capable to modify the geometry of our space-time.

 We think that this example might offer new interesting possibilities for constructing new models of theories with extra dimensions and about the relative dimensional reduction.
 
 About this point, it will be interesting to study the extension of this model to $p$-branes. 
 Using the pull-back of the MC forms in eq. \eq{1.4}
we can construct an action invariant under the coadjoint Poincar\'e algebra for strings and more generally for 
 $p$-branes. For instance we can  consider the string extension of the diff-invariant model for a point particle considered previously.  We will describe the strings in 
 the Polyakov formulation with a metric in the world-sheet \cite{Zwiebach:2004tj}
 \footnote{we will use the notations of  Zwiebach's book \cite{Zwiebach:2004tj}}.
 All the dynamical variables are functions of the two variables $(\sigma,\tau)$ which are also denoted by $\xi^\alpha,~\alpha=1,2$. We will introduce also the following quantities
\be
\gamma^1_{\alpha\beta} =\frac{\de x^\mu}{\de \xi^\alpha}\frac{\de x_\mu}{\de \xi^\beta},~~~
\gamma^2_{\alpha\beta} =\frac 12\frac{\de \xi^{\mu\nu}}{\de \xi^\alpha}\frac{\de \xi_{\mu\nu}}{\de \xi^\beta},~~~
\gamma^3_{\alpha\beta} =\left(\frac{\de \eta^\mu}{\de\xi^\alpha}-\xi^{\mu\rho}\frac{\de  x_\rho}{\de \xi^\alpha}\right)
\left(\frac{\de \eta_\mu}{\de\xi^\alpha}-\xi_{\mu\sigma}\frac{\de  x^\sigma}{\de \xi^\alpha}\right)
\ee
Then, the Polyakov action analogous to the one that we have used for the point particle is
\be
S=-\frac 1{4\pi\alpha'}\int d\sigma d\tau\, \sqrt{-h} h^{\alpha\beta} (
\gamma^1_{\alpha\beta}+z\gamma^2_{\alpha\beta} )-\frac 1{4\pi\beta'}\int d\sigma d\tau\, \sqrt{-h} h^{\alpha\beta}\gamma^3_{\alpha\beta}
\ee
where $h^{\alpha\beta}$ is an arbitrary two-dimensional metric, $h={\rm det}(h_{\alpha\beta})$. Eliminating the zweibeins via their equations of motion, one gets:
\be 
S=-\frac 1{2\pi\alpha'}\int d\sigma d\tau\, \sqrt{{-\,\rm det} (
\gamma^1_{\alpha\beta}+z\gamma^2_{\alpha\beta} )}-\frac 1{2\pi\beta'}\int d\sigma d\tau\,\sqrt{-{\rm det}(\gamma^3_{\alpha\beta})}
\ee
This action can be easily generalised to the action of  a $p$-brane.  For this, it  is sufficient to go from a 2-dimensional world-sheet to a $p+1$ dimensional, implying $\alpha,\beta = 1,\cdots,p+1$ and to replace the volume element
\be
d\sigma d\tau \to d\xi_1\cdots d\xi_{p+1}
\ee

Also in this case, neglecting the second term in the action, we obtain the action for a $D(D+1)/2$ dimensional string (or $p$-brane). In the case of $D=4$ a 10-dimensional  extended object.

 \section*{Acknowledgments}
 We acknowledge discussions with Axel Kleinschmidt and Patricio Salgado-Rebolledo.
JG has been supported in part by MINECO FPA2016-76005-C2-1-P and Consolider
CPAN, and by the Spanish government (MINECO/FEDER) under project MDM-2014-0369
of ICCUB (Unidad de Excelencia María de Maeztu).

\end{document}